%
%
%
%
\input harvmac.tex
\def\Titlemz#1#2{\nopagenumbers\abstractfont\hsize=\hstitle\rightline{#1}%
\vskip 0.4in\centerline{\titlefont #2}\abstractfont\vskip 0.3in\pageno=0}
%
\Titlemz{\vbox{\baselineskip 12pt \hbox{LAVAL-PHY-22-96}}}
{\vbox {\centerline{ The structure of conserved charges in open spin chains} }}
\centerline{M. P. Grabowski and P. Mathieu$^*$}
\smallskip
\centerline{ \it D\'epartement de
Physique, Universit\'e Laval, Qu\'ebec, Canada G1K 7P4}
\vskip.4in
\bigskip
\centerline{\bf Abstract}
\bigskip\vskip 0.45cm
We study the local conserved  charges in integrable spin chains
of the XYZ type
with nontrivial boundary conditions. The general structure of these
charges consists of a bulk part, whose density is identical to that
of a periodic chain, and  a boundary part.
In contrast with the periodic case, only the charges corresponding to
the interactions of even number of spins exist for the open chain.
Hence, there are half as many charges in the open case as in the
closed case.
For the open spin-1/2 XY chain, we derive the explicit expressions of all
the charges. For the open spin-1/2 XXX chain,  several lowest order charges are
presented and a general method of obtaining the boundary terms is indicated.
In contrast with the closed case, the XXX charges cannot be described
in terms of a Catalan tree pattern.

\bigskip\vskip 0.5cm

\vskip 1.2 in
{\noindent $^*$ Work supported by NSERC (Canada) and FCAR (Qu\'ebec).  }
\noindent\smallskip
\Date{04/96}

\vfill\eject

\newcount\eqnum \eqnum=1
\def\eq{
\eqno(\secsym\the\meqno)
\global\advance\meqno by1
 }
\def\eqlabel#1{
{\xdef#1{\secsym\the\meqno}}
\eq
}

\newwrite\refs
\def\startreferences{
 \immediate\openout\refs=references
 \immediate\write\refs{\baselineskip=14pt \parindent=16pt \parskip=2pt}
}
\startreferences

\refno=0
\def\aref#1{\global\advance\refno by1
 \immediate\write\refs{\noexpand\item{\the\refno.}#1\hfil\par}}
\def\ref#1{\aref{#1}\the\refno}
\def\refname#1{\xdef#1{\the\refno}}

\let\rw\rightarrow
\def\tr{{\rm tr}}
\def\C{{\cal C}}
\def\R{{\cal R}}
\def\P{{\cal P}}
\def \S{{\bf {\sigma} }}
\def \s{{ {\sigma} }}
\def \bul{{\bullet\;}}
\def \cir{{\circ\;}}
\def \a{{\lambda_x}}
\def \b{{\lambda_y}}
\def \c{{\lambda_z}}
\def \frac#1#2{{ \textstyle{#1 \over #2} }}

\def\t{\tau}

\def\ua{\uparrow}
\def\da{\downarrow}

\def\dg{\dagger}

\def\text#1{\quad\hbox{#1}\quad}


\def\ubrackfill#1{$\mathsurround=0pt
        \kern2.5pt\vrule depth#1\leaders\hrule\hfill\vrule depth#1\kern2.5pt$}
\def\contract#1{\mathop{\vbox{\ialign{##\crcr\noalign{\kern3pt}
        \ubrackfill{4pt}\crcr\noalign{\kern3pt\nointerlineskip}
        $\hfil\displaystyle{#1}\hfil$\crcr}}}\limits
}

\def\ubrack#1{$\mathsurround=0pt
        \vrule depth#1\leaders\hrule\hfill\vrule depth#1$}
\def\dbrack#1{$\mathsurround=0pt
        \vrule height#1\leaders\hrule\hfill\vrule height#1$}
\def\ucontract#1#2{\mathop{\vbox{\ialign{##\crcr\noalign{\kern 4pt}
        \ubrack{#2}\crcr\noalign{\kern 4pt\nointerlineskip}
        $\hskip #1\relax$\crcr}}}\limits
}
\def\dcontract#1#2{\mathop{\vbox{\ialign{##\crcr
        $\hskip #1\relax$\crcr\noalign{\kern0pt}
        \dbrack{#2}\crcr\noalign{\kern0pt\nointerlineskip}
        }}}\limits
}


\newsec{Introduction}
Recently we have obtained the structure of the local conserved charges for the
XXX
chain [\ref{Grabowski M P  and  Mathieu P {\it Ann. Phys.} {\bf 243}
(1995) 299}\refname\Ann]
and some of its generalizations: the isotropic $su(N)$ spin chain and the
octonionic
model [\ref{Grabowski M P and Mathieu P {\it J. Math. Phys}.
{\bf 36} (1995) 5340}\refname\JMP].
  Explicit expressions -- in the form of a Catalan tree pattern -- have
been derived for all the charges.   Different extensions of
these results can be considered. In the present work we study
the modification of the Catalan tree pattern for open
 (finite) chains.\foot{As far as the pattern of charges
is concerned, the infinite open chain is not distinguishable  from the closed
case.}

As for closed chains, there is a transfer
matrix formalism for open chains from which the conserved quantities can be
obtained by
power expansion in terms of the spectral parameter.  For closed  chains, this
approach is not an effective way of
deriving  the explicit expressions of the conserved charges.  In
the open case, the situation  is even worse
as the open transfer matrix is, roughly speaking,
 the trace of the square of the closed chain monodromy matrix
 (up to boundary terms).  A somewhat surprising consequence of this squaring
process is that half of the local conserved charges of the  closed chain
disappear when
the chain is cut open.
More precisely, as we will show below,
the expression for the open conserved charges
contain a bulk part and a boundary part.  In the infinite chain limit, the
boundary
part is irrelevant: the bulk part is thus necessarily the full closed chain
charge
expression.  Hence, a charge exists in the open case only if
suitable  boundary terms can be added to the bulk part to preserve the
commutativity of the total charge with the hamiltonian.

It turns out that conservation preserving boundary terms do not always exist.
For instance, for the XY closed chain, there are
two infinite sequences of local charges,
that is,
$\{H_n^{(1)}\}$ and
$\{H_n^{(2)}\}$, for all integer values of
$n$, where $n$ indicates the highest number of adjacent spin factors of the
charge density.  In the open case, there are no appropriate boundary
counterterms for half of these charges. There
remains only one charge for each value of $n$.  For the closed XXX chain, we
know that
there exists one local charge $H_n$ for each positive integer $n$.  However,
for
the open XXX case, only the charges with $n$ even survive.
In particular,  $H_3$ does not exist!


A major technical difficulty encountered in deriving explicit expressions for
the
conserved charges for open chains is caused by the non-existence of a
boost operator, whose commutation with the hamiltonian would generate the
higher order
charges.  As a result, there is no systematic and effective way of obtaining
the
expressions for the first few charges.  In spite of this, one could try to
find them by brute
force and see whether an underlying pattern emerges.  From the
expressions of $H_4$ and $H_6$ obtained for the XXX open chain, this would seem
to be so: their
boundary parts can be written in terms of the same spin polynomials as their
bulk
part and they can actually be expressed in terms of lower order `bulk charges'
localized at the boundaries.  However, this is not a generic feature: $H_8$
cannot be
written in this polynomial basis.

The article is organized as follows.  In the next section, we review the
transfer
matrix formalism.  This is used to show the absence
of odd degree local conserved charges.  A second argument is worked out in the
following
section, from a computational approach: the charge is written as a sum of a
(known)
bulk piece and a set of boundary counterterms to be fixed.  We  show that in
$H_n$ the counterterms must cancel a Catalan-tree bulk-type charge localized at
each
boundaries.  We then demonstrate that no such terms can be found for
$H_3$ and then for all
$H_{2n+1}$.  The explicit form of the first few nontrivial charges is
presented.  Unfortunately the simple  polynomial basis used to describe
the closed XXX charge is inadequate in the open case, thus making it
very hard to find a general expression for all the charges.
Straightforward generalizations and conclusions
are reported in the final section.

\newsec{The general structure of local conservation laws in open chains}

In this section we briefly recall  the transfer matrix formalism and use it
to determine the general structure of conservation laws in open spin chains.
This will be done first for the relatively simple spin-1/2 XXX model, and
later the extension of the argument to
more general integrable chains will be indicated.

The integrable spin chain with nontrivial boundary
conditions [\ref{Sklyanin E K {\it J. Phys. A: Math. Gen.} {\bf 21}
(1988) 2375}]
is described by the transfer matrix
$$t_o(u) = \tr(K_+(u)T(u)K_-(u)T^{-1}(-u)).  \eqlabel\tmodef$$
$u$ is the spectral parameter and  $T(u)$ is the monodromy matrix,
whose trace gives the usual (closed) chain transfer matrix:
$$t_c(u)=\tr\, T(u)\eqlabel\tc$$
(the subscripts `o' and `c' refer to open and closed cases respectively).
The monodromy matrix is constructed from  the basic
$R$-matrix of the model
$$T(u)= R_{N0}\cdots R_{10}.\eq$$
The index $i=1, \cdots,N$ labels a vector space  at site number $i$ and $0$
refers
to internal or auxiliary
space, over which the trace in (\tmodef) is taken.
$K_+, \,K_-$ are respectively  left and right boundary
matrices.
For the XYZ model,
the most general form of these matrices compatible with
integrability has been derived in [\ref{Inami T, Konno T {\it J. Phys. A: Math.
Gen.} {\bf 27} (1994) L913}].   For free open boundaries, to which we
confine ourselves for the most part of this work, $K_\pm = I$, the $2\times
2$ identity matrix.

The transfer matrix can be used as a generating function for the conserved
quantities. It follows from the Yang-Baxter equation that any derivative of
the transfer matrix commutes with the hamiltonian (which is, by construction,
the first logarithmic derivative of the transfer matrix, taken at the
 particular value $u=0$ of the spectral parameter).
The quantities obtained in this way are however nonlocal, i.e.,
they contain spin interactions at distances growing  arbitrarily when the
length of the chain is increased.
To obtain local conserved quantities out of the transfer matrix, one needs
to take logarithmic derivatives [\ref{L\"uscher M {\it Nucl. Phys.} {\bf B117}
 (1976) 475}].
The argument of L\"uscher, showing
the local nature of the logarithmic derivatives of $t_c(u)$ at $u=0$,
remains valid for open chains.
However, there are two new features in the the open case as
compared to the periodic one.
First, translational invariance is obviously lost.
Second, and more surprisingly, all charges
with an odd number of interacting spins disappear!

We will first concentrate on the spin-$1/2$ XXX chain, whose $R$ matrix takes
the simple form
$$R= uI +  P\eq$$
where $P$ is the permutation operator.
The derivatives of the transfer matrix of the closed chain are:
$$ t_{c}^{(n)} (0) \equiv {d^n t_c(u)\over {du^n}}\Big|_{u=0}= n!
\sum_{| \C| =N-n} {\cal P}_R(\C), \eqlabel\tmcl$$
where the sum goes over all ordered clusters $\C$ of ($N-n$) points,
that is, sets $\{i_1, \dots i_{N-n}\}$ with $i_1<i_2<\dots<i_{N-n}$.
${\cal P}_R(\cal C)$ (and ${\cal P}_L(\C) $ appearing below)
denote  the  cyclic permutation
of the spins of the sequence $\C$ to the right (or to the left, respectively).
In particular, $$ t_{c}(0) =
{\cal P}_R(\Lambda),\eq$$
where $\Lambda=\{1, 2, \dots, N\}$.
The logarithmic derivatives of the transfer matrix yield local
charges. In particular, using
$$t_{c}^{-1}(0) =
{\cal P}_L(\Lambda), \eq$$
it  can be easily seen that
$$ t_{c}^{-1}(0) t_c^{(1)}(0)=
\sum_{i=1}^N {\cal P}_R(\{i,i+1\})=\sum_{i=1}^N P_{i,i+1}.\eqlabel\clham $$
The permutation operator above can be written as
$$P_{i,i+1}=\frac{1}{2} (1+ \S_i \cdot\S_{i+1}) \eq$$
(where
$\s_i$ denotes a Pauli matrix at site $i$) and thus
(\clham), up to a constant,  coincides with the hamiltonian of the periodic XXX
chain.

Let $Q_n^c$ denote
the $(n-1)$-th logarithmic derivative of $t_c(u)$ at $u=0$:
$$ Q_n^c= {d^{n-1}\over {du^{n-1}} } \ln t_c(u)|_{u=0}.\eqlabel\Qndef $$
In terms of $v_k$ defined as
$$v_k=T^{-1}(0) T^{(k)}(0) \eq$$
we have
$$ Q_{n}^c= \tr\, v_{n-1} + \tr\, p_{n-1}(v_1, \dots, v_{n-2}), \eqlabel\qncl$$
where
$p_{n-1}$ is a homogeneous polynomial (of degree $n-1$) in $v_1,\dots,
v_{n-2}$.
These charges are translationally invariant and can be written in the form
$$ Q_n^c=\sum_{j=1}^{N} q_n(j) \eqlabel\qnden$$
where the density $q_n(j)$ contains interactions of at most $n$ spins
on sites $\{j, j+1, \dots, j+n-1\}$ (addition understood modulo $N$).

Consider now the chain with open free boundaries ($K_\pm=I$).
Conserved charges may be obtained from the
expansion of the logarithm of the transfer matrix $t_o(u)$:
$$ Q_n^o= {d^{n-1}\over {du^{n-1}} } \ln t_o(u)|_{u=0}.\eqlabel\Qndefop $$
We will now argue that for $n$ even, the leading term
-- that with the greatest number of interacting spins -- in the above
expression is the same as in the closed case.
Taking the trace over the two-dimensional
	     auxiliary space, we obtain
$$t_o(0)= 2 I\eq$$ where $I$ is the
$2^N \times 2^N $ identity matrix.  Therefore, $t_o^{-1}(0)$ is
	    simply 1/2, and
$$ Q_{n}^o= \frac{1}{2} t_o^{(n-1)}(0) + p_{n-1}
(\frac{1}{2} t_o^{(1)}(0), \dots, \frac{1}{2}  t_o^{(n-1)}(0) ) \eqlabel\qnop$$
where $p_{n-1}$ is the same polynomial as in (\qncl).
Consider first $n$ to be even.
{}From (\tmodef)  we get
$$ t_o^{(n-1)}(0) = 2\; \tr\, v_{n-1}  \;  + \tr\, r_{n-1} \eq$$
where $r_{n-1}$ is
a homogeneous polynomial of degree $n-1$ in the variables $v_1,\dots, v_{n-2}$.
Note that
(\qnop) and (\qncl) both contain the same local term
$$\tr\, v_{n-1} \sim  \sum_i  {\cal P}_R(\{i, i+1,\dots, i+n-2\}) \eq$$
(with the addition understood modulo $N$ for the closed chain).
For the closed XXX chain, the set of local conserved charges
obtained from the transfer matrix has been shown to be complete
 (see [\ref{Babbit D and Thomas L {\it J. Math.  Anal.} {\bf 72}
(1979), 305}\refname\Babbi]).
This means that any local conserved quantity in the closed chain
must be a linear combination of the logarithmic derivatives of the transfer
matrix
 $t_c(u)$.
In particular, this implies that given the leading term in $Q_n^c$ (the one
describing $n$-spin interactions), there is a unique set of subleading terms
ensuring the commutativity of $Q_n^c$ with the hamiltonian.
The introduction of  nontrivial boundary conditions can be viewed as
a perturbation of the original periodic chain
\foot{If $H_c$ denotes the hamiltonian of a periodic spin chain with
nearest-neighbor interactions, $H_c=\sum_{j=1}^N h_{j,j+1}$,
by adding the single perturbation term $-h_{1,N}$ one
obtains the open-ended chain $H_o=H_c - h_{1,N}$.} and this can only
reduce the number of conservation laws. The key point is that,
as can be seen from (\qnop) and (\qncl) for $n$ even, the leading term of a
charge $Q_n^o$  must be the same as for the closed chain.
Consider now the infinite limit of the open chain. Obviously, in this limit
(\Qndef)
must coincide with some function of conserved quantities for the infinite
periodic chain.
But the only local conserved quantities in this case are
linear combinations of the charges $ Q_n^c$. Therefore, the bulk density
of the open chain $n$-spin charge $Q_n^o$ must be the same as the density of
$Q_n^c$.
It follows that for $n=2m$,  $Q^o_{2m}$ has the form
$$Q^o_{2m}= Q_{2m}^{bulk} + q_{2m}^L +q_{2m}^R ,\eqlabel\chargeform$$
where
$$ Q_{2m}^{bulk}= \sum_{j=1}^{N-2m+1} q_{2m}(j) ,\eq$$
(i.e., we take only clusters that do not `jump' over the boundary)
and
$q_{2m}^L$ ($q_{2m}^R$) stands for boundary terms at the left (right)
boundary. As we will show in sect. 3, the boundary terms involve at most
the $2m-2$ spins
adjoining the boundary.

In contrast, when $n$ is odd, i.e.,
$n=2m+1$, there is no totally local term in $Q^o_n$:
$$ Q^o_{2m+1}=p_{2m}(\frac{1}{2} t_o^{(1)}   , \dots,  \frac{1}{2} t_o^{(n-1)})
\eq$$
and the  $n$-spin  contribution vanishes.
In consequence, the bulk part of $Q^o_{2m+1}$
is zero,  implying  that for $n$ odd, $Q^o_n$ must be trivial.

The absence of odd-order charges
is in fact a simple consequence of the chain reversal symmetry
of the open chain, i.e., its invariance under the transformation
$${\S}_n \to {\S}_{N-n+1}. \eqlabel\chainrev$$
This symmetry is specific to the open case: the transfer
matrix of the periodic XXX chain is not invariant under (\chainrev):
$$ \tr (R_{N,0}(u)R_{N-1,0}(u)\dots R_{1,0}(u) ) \neq \tr (R_{1,0}(u)
R_{2,0}(u)
\dots R_{N,0}(u)) .\eq$$
Indeed, it is easy to see from the explicit form of the local
conserved quantities of the XXX chain (see [\Ann]),
that the logarithmic derivatives
of $t_{c}(u)$ of odd degrees
(corresponding to charges with an even number of spins)
do not change under (\chainrev), while those of even degree
(corresponding to odd number of spins) change sign.

In contrast,
the transfer matrix of the open XXX chain is invariant under the
 operation (\chainrev).
Using
$$ R_{n,0}^{-1}(-u)=- {1 \over {1+u^2} }R_{n,0} \eq$$
$t_o(u)$ can be written in the form
$$ \eqalign{t_{o}(u)= & \tr (R_{N,0}\dots R_{1,0}(u)R_{1,0}^{-1}(-u)\dots
R^{-1}_{N,0}(-u)
\cr = &{(-1)^N \over {(1+u^2)^N }}
\tr ( R_{N,0}(u)\dots R_{2,0}(u) R_{1,0}^2(u) R_{2,0}(u)\dots R_{N,0}(u) ).}
\eqlabel\trmb$$
 Under (\chainrev), it transforms as
$$ t_o(u) \to {(-1)^N \over {(1+u^2)^N }}
\tr ( R_{1,0}(u)\dots R_{N-1,0}(u) R_{N,0}^2(u) R_{N-1,0}(u)\dots R_{1,0}(u) )
\eq$$
which, due to the cyclic property of the trace, is equal to (\trmb).
This symmetry excludes thus the possibility of building-up the
bulk density of $Q_{2m+1}^o$ from the
densities of odd-spin charges $Q_{2m+1}^c$
since their sign is changed  under (\chainrev).
Hence, there are no odd-spin charges for the
open XXX chain.

The above argument for the spin-1/2 XXX chain may be similarly
applied to the open XYZ
chain.
As before, the key step is to observe that, for $n$ even,
(\qnop) and (\qncl) both contain the same local leading term with
$n$ spins interacting.
In the XXX case, one could use the  completeness property of the system
of charges generated by the transfer matrix of the periodic chain
to show that the bulk density in the open case must coincide with
the density of the periodic chain.
To our knowledge, the
completeness of the system of charges (\Qndef) for the general XYZ case
has not been established.
However, considering the infinite limit of the open chain, one may conclude
that
the bulk density of
the conserved charge (\Qndef) must coincide with the density of the
corresponding closed chain expression, modulo the
densities of additional conserved charges
(which do not have to be generated by the transfer matrix (\tc)).
Therefore, the set of conserved quantities generated from the transfer matrix
still has the general structure  (\chargeform), but the bulk density
may now, a priori, contain additional contributions.
For $n$ odd,  there is no $n$-spin local term in (\qnop), and in
consequence the bulk density of $Q_{2m+1}$ must vanish.
As for the XXX case, this can be viewed as a
consequence of the chain reversal symmetry (\chainrev).
Recall that the basic $R$ matrix of the XYZ model
satisfies
the unitarity requirement
$$ R(u) R(-u) =\rho (u)  I  \eqlabel\runit$$
where $\rho(u)$ is some scalar function.
Using (\runit), the transfer matrix of the
open XYZ chain can be rewritten in a form which is manifestly invariant under
(\chainrev):
$$ \eqalign{t_{o}(u)= &
\tr ( R_{N,0}(u)\dots R_{2,0}(u) R_{1,0}(u) R^{-1}_{1,0}(-u)
R^{-1}_{2,0}(-u)\dots R^{-1}_{N,0}(-u) ) \cr=& {1 \over {\rho(u)^N }}
\tr ( R_{1,0}(u)\dots R_{N-1,0}(u) R_{N,0}^2(u) R_{N-1,0}(u)\dots R_{1,0}(u) )
.}
\eq$$
Note that the above argument does not exclude a possible
existence of  additional odd-spin conserved charges
not given by the formula (\Qndef).
Such a situation indeed exists for the
XY model (see sect. 4). However, this is not expected in the non-degenerate XYZ
case, for which the system of charges generated from the transfer matrix is
likely to be complete.

\newsec{The calculation of the conserved charges for the open XXX model}

The model-defining hamiltonian for the XXX chain with
free open boundaries is\foot{
{}From now on we omit the superscript `o' from conserved charges.}
$$H_2 = \sum_{i=1}^N \S_i\cdot\S_{i+1}\eq$$
where $\S$ stands for the vector $(\sigma^x, \sigma^y, \sigma^z)$,
$\sigma$'s being the usual Pauli matrices.
As we have seen in the previous section,
higher-order charges will have a bulk part and a boundary part. It is
convenient
to redefine the charges so that in the bulk they will be described by the
Catalan tree pattern [\Ann]. (This corresponds to a change of basis from
$\{Q_n\}$ to $\{H_n\}$  where
such that $H_n$ does not contain any linear combination of lower order
$H_m$'s. This change of basis corresponds to taking linear
combinations of the logarithmic derivatives of the transfer matrix).
We thus look for charges in the form
$$H_n= H_n^{{\rm~bulk}} +h_n^{{\rm~L}}+h_n^{{\rm~R}},\eq$$
where $h_n^{{\rm~L}}$ and
$h_n^{{\rm~R}}$ are boundary terms located respectively at
the left and right extremity of the chains (their precise extension will be
evaluated later);  these are
the unknowns to be determined.  With the above ansatz, we assume the chain
to be sufficiently long  to prevent a mixing of boundary terms from
opposite boundaries.
Recall that the local integrals of
motion for the closed XXX model can  be expressed as linear combinations of the
$F^c_{n,k}$'s defined as
$$ F^c_{n,k}=\sum_{\C \in \C^{(n,k)}} f_n(\C)\eq$$
$\C^{(n,k)}$ stands for the set of all $n$
ordered sites with $k$ holes (i.e., the  sites are not necessarily
adjacent) and $\C$ is some cluster among this set.
$f_n(i_1,i_2,\cdots,i_n)$ is defined as
$$\eqalign{
& f_0=f_1=0,\cr
&  f_2 = \S_{i_1}\cdot \S_{i_2},\cr
& f_3 = (\S_{i_1}\times \S_{i_2})\cdot\S_{i_3},\cr
&\cdots\cr
& f_n = (\cdots((\S_{i_1}\times \S_{i_2})\times\cdots)\cdot\S_{i_n}. \cr}
\eq$$
(Notice that the parentheses are nested toward the left).
The closed chain conservation laws  $H_n^c$'s are
$$ H_n^c= F^c_{n,0}+ \sum _{k=1}^{[n/2]-1} \sum_{\ell=1}^{k}
\alpha_{k,\ell} F^c_{n-2k,\ell}\eqlabel\hncl$$
where
the coefficients $\alpha_{k,\ell}$ are
the generalized Catalan numbers
\eqn\Catalrel{ \alpha_{k,\ell}=\pmatrix {2k-\ell \cr k-\ell+1}-
\pmatrix {2k -\ell \cr k-\ell-1}.}
The bulk part of an open-chain conserved charge (of even order) is then
given simply by restricting the corresponding closed chain expression to those
clusters which do not `jump over the border'.
More precisely, let ${\cal O}^{(n,k)}$ denote the subset of $\C^{(n,k)}$
obtained by
removing the clusters which overlap the borders.
(For instance, for $N=6,\;
{\cal O}^{(3,3)}=(1,2,6), (1,3,6), (1,4,6),(1,5,6)$.)
Denoting $$ F^o_{n,k}=\sum_{\C \in {\cal O}^{(n,k)}} f_n(\C),\eq$$
$H_n^{bulk}$ is given by formula (\hncl) with $F_{n,k}^c$ replaced by
$F_{n,k}^o$.

For the purpose of determining the boundary terms, we can focus on a single
extremity,
or equivalently, consider a semi-infinite chain ($N\rw\infty$, so that
$i=1,2,\cdots$). The term
$h_n^{{\rm~L}}$ can then be characterized as follows.  It is designed to
cancel,
in the commutator $[H_2,H_n]$, those terms that would be cancelled for an
infinite
chain ($i\in{\bf Z}$) by the contribution of the `link'  $\S_0\cdot\S_1$. In
the infinite chain commutator $[\S_0\cdot\S_1,H_n^{(\infty)}]$, these are
precisely those
resulting terms that live in the semi-infinite chain, i.e., on sites $i\geq
1$.  Our first step is thus to evaluate this commutator:
$$[ H_2, H_n^{{\rm~bulk}}] = -[ \S_0\cdot\S_1,
H_n^{(\infty)}]\Big|_{i\geq1}\equiv
\R_n^L \eqlabel\catatbor$$ where $\R_n^L$ stands for left remainder.  This
expression is
readily evaluated: it is a sort of anti-boost, that is, the inverse of the
boost action
that generates
$H_{n+1}$ from
$H_n$. In the boost operation, $[H_2, H_n]$ gives, in addition to $H_{n+1}$,
some
lower order local charges. This turns out not to be the case for the anti-boost
calculation.  We find
the  remarkably simple result
\eqn\reste{[ \S_0\cdot\S_1, H_n^{(\infty)}]\big|_{i\geq1}
 =
4i\;H_{n-1}^{{\rm~bulk}}\Big|_{{\rm~L-bdry}} }
where $H_{n-1}^{{\rm~bulk}}$ restricted to the left boundary contains only
those terms
of
$H_{n-1}^{{\rm~bulk}}$ whose leftmost spin is at site 1.
This is proved below.

Let us first introduce a schematic way of describing the related calculations.
To represent
the monomial $f_n(i_1,i_2,\cdots,i_n)$
we use a
sequence of dots, with black dots corresponding to occupied sites.
For our purposes, we will only need sequences  of dots based
at site 0 (that is, the first dot represents $i_1=0$).
 For instance
$f_6(0,...,5)$ and $f_4(0,1,3,5)$ are represented respectively by
$$f_6(0,...,5) =
((((\S_0\times\S_1)\times\S_2)\times\S_3)\times\S_4)\cdot\S_5\sim
\bullet\;\bullet\;\bullet\;\bullet\;\bullet\;\bullet\; \eq$$
and
$$f_4(0,1,3,5)= ((\S_0\times\S_1)\times\S_3)\cdot\S_5\sim
\bullet\;\bullet\;\circ\;\bullet\;\circ\;\bullet\eq$$
The commutation relation
$$[\S_0\cdot\S_1, (\S_0\times\S_1)\cdot\S_2] = 4i(\S_1\cdot\S_2 -
\S_0\cdot\S_2)\eq$$
will then
be represented by the following diagram:
$$ \contract{\bul\bullet}\;\bul = 4i\; \cir\bul\bul \; - \; 4i\;  \bul\cir\bul
\eqlabel\diagone$$
  The contraction in this diagram indicates taking the value
of the commutator of the `link' (i.e., the $\S_0 \cdot \S_1$ operator)
with the $f$ polynomial represented by a sequence of dots.
In the following, we
will implement the projection onto the semi-infinite chain by setting to zero
all spin
factors at site $i\leq0$, that is, by
droping the terms whose first dot is occupied.
This projection will be indicated by an arrow. For example, in (\diagone)
it amounts to setting to zero the term $\S_0\cdot \S_2$:
$$ \contract{\bul\bullet}\;\bul \to  4i\; \cir\bul\bul \;  \eq$$
Here
are sample calculations of $[ \S_0\cdot\S_1, H_n^{(\infty)}]\big|_{i\geq1}$.
For $n=4$,
$$ H_4^{(\infty)} = F_{4,0}+F_{2,1} \eq$$
It is clear that the only terms in $H_4$ that will
contribute to this commutator, after the projection, are those whose leftmost
spin is at
site 0.  The contribution of $F_{4,0}$ is
$$ \contract{\bul\bullet}\;\bul\bul  \to  4i\; \cir\bul\bul\bul \; . \eq$$
The  commutator with $F_{2,1} $ does not survive the projection:
$$ \contract{\bul\circ}\;\bul =   2i\; \bul \cir\bul \; \to 0 \;  \eq$$
The result is exactly the part of $F_{3,0}$ based at
the left boundary, that is,
$$\R_4^L =-4 i H_{3}^{{\rm~bulk}}\Big|_{{\rm~L-bdry}} . \eq$$
For $n=5$,
$$ H_5^{(\infty)} = F_{5,0}+F_{3,1} \eq$$
we have
$$ \contract{\bul\bullet}\;\bul\bul\bul \to   4i\;  \cir\bul\bul\bul\bul  \;
\eq$$
and
$$ \contract{\bul\bullet}\;\cir\bul \to   4i\;  \cir\bul\cir\bul  \;  \eq$$
We thus obtain
$$\R_5^L =-4 i H_{4}^{{\rm~bulk}}\Big|_{{\rm~L-bdry}}.\eq$$

Having worked out these examples,
let us return to the proof of the general identity
\reste. The contribution of a term $F_{n,k}$ to the commutator is clearly
$4i\; F_{n-1,k}\,$ restricted to the left boundary: indeed, only monomials of
the form
$f_{n}(0,1,\cdots,n-1)$ contribute in the commutator and after commutation and
projection, each such term is transformed into $4 i\; f_{n-1}(1,\cdots,n-1)$,
which builds
up $F_{n-1,k}$ restricted to clusters starting at site 1.  (The number of holes
is
unaffected in this process.) If
$n$ is odd, the structure of the Catalan tree is not modified and
$H_n^{(\infty)}$ is mapped to
$4i\; H_{n-1}^{{\rm~bulk}}\Big|_{{\rm~L-bdry}}$.  If
$n$ is even, the terms $F_{2,k}$ do not contribute and we simply strip off the
last
row of the Catalan tree of $H_n^{(\infty)}$ to get the tree of
$H_{n-1}^{(\infty)}$,
restricted to the left boundary.

Up to this point, we have only found what the boundary terms $h_n^{{\rm~L}}$
have
to cancel when commuted with $H_2$: we require
$$[H_2, h_n^{{\rm~L}}] =
-\R_n^L = 4i\; H_{n-1}^{{\rm~bulk}}\Big|_{{\rm~L-bdry}}. \eqlabel\basicl$$
The existence of a conserved charge is thus reduced to the existence a
suitable $h_n^{{\rm~L}}$ satisfying this equation.

In this framework, the non-existence of $H_3$ is simply established: there
are no $h_3^{{\rm~L}}$, build up with 1 or 2 spin factors, that could cancel
$4i\; \S_1\cdot\S_2$ upon commutation with
$\S_1\cdot\S_2+ \S_2\cdot\S_3$.  The same argument also implies the
non-existence of
any odd charge, simply because $H_{2n+1}^{(\infty)}$ contains a three-spin
piece and
$h_{2n+1}^{{\rm~L}}$ would have to contain exactly the same counterterm
required
to make a charge $H_3$.

Equation (\basicl) implies that the boundary part in $H_n$ may contain only
interactions of up to $n-2$ spins adjoining the boundary. For the spin-1/2
chain,  the most general term containing multi-spin interactions is
a multilinear polynomial in spin variables, which  can be
equivalently represented in the permutation basis as a linear combination of
permutations of the $n-1$ spins at the boundary (with $(n-1)! $ arbitrary
coefficients). These coefficients
are then fixed by enforcing the condition (\basicl).
The first two boundary terms are:
$$h_4^L= 4 {\cal P}_L({1,2}),\eq$$
$$\eqalign{
h_6^L=&  4(- \P_L(1,2,3,4) - \P_L(4,3,2,1) + \P_L(1,2,4,3)+ \P_L(2,1,3,4))\cr
&+ 8 {\cal P}_R (1,3) - 8 {\cal P}_R(2,3).}
\eq$$
A particularly useful  set of polynomials in spin variables is provided by
the $f_n(\C)$'s (where $\C$ is an ordered cluster ($i_1<\cdots i_n$)),
introduced above as the building blocks of the bulk
part of $H_n$.  They can be expressed in terms of permutations as
follows:
$$ f_n(\{ i_1, i_2, \dots, i_n\})=
2 (-i)^{n-2}[\dots
[[ P_{i_ni_{n-1}}, P_{i_{n-1}i_{n-2}}], P_{i_{n-2}i_{n-3}}]\dots, P_{i_2i_1}]
\eq$$
with
$$ f_2(i_1,i_2)= 2 P_{i_1i_2} -1. \eq$$
In particular,
$$ \eqalign{
f_3(i_1,i_2, i_3)=& 2 i ( \P_L(i_1, i_2, i_3)-\P_L(i_3, i_2, i_1)), \cr
f_4(i_1,i_2, i_3, i_4)=& - 2 (
\P_L(i_1,i_2,i_3,i_4) +\P_L(i_4,i_3,i_2,i_1)  \cr &
- \P_L(i_1,i_2,i_4,i_3)  - \P_L(i_3,i_4, i_2, i_1)), \cr
f_6(i_1,i_2, i_3, i_4, i_5, i_6)= & 2 (
\P_L(i_1, i_2, i_3, i_4, i_5, i_6)  - \P_L(i_1, i_2, i_3, i_4, i_6, i_5)
\cr &
- \P_L(i_1, i_2, i_3, i_5, i_6, i_4) + \P_L(i_1, i_2, i_3, i_6, i_5, i_4)
\cr &
 - \P_L(i_1, i_2, i_4, i_5, i_6, i_3) + \P_L(i_1, i_2, i_4, i_6, i_5, i_3)
\cr &
+ \P_L(i_1, i_2, i_5, i_6, i_4, i_3) - \P_L(i_1, i_2, i_6, i_5, i_4, i_3)
\cr &
- \P_L(i_1, i_3, i_4, i_5, i_6, i_2) + \P_L(i_1, i_3, i_4, i_6, i_5, i_2)
\cr &
+ \P_L(i_1, i_3, i_5, i_6, i_4, i_2) - \P_L(i_1, i_3, i_6, i_5, i_4, i_2)
\cr &
+ \P_L(i_1, i_4, i_5, i_6, i_3, i_2) - \P_L(i_1, i_4, i_6, i_5, i_3, i_2)
\cr &
 - \P_L(i_1, i_5, i_6, i_4, i_3, i_2) + \P_L(i_1, i_6, i_5, i_4, i_3, i_2) ).
\cr
} \eq$$

The boundary terms $h_4^L$ and $h_6^L$ have a simple form when
expressed in the $f_n({\cal C})$ basis.
Define $$ f_{n,k}(j)=\sum_{\C\in {\cal {C}}^{(n,k)}(j)}f_n(\C) \eq$$
where ${\cal C}^{(n,k)}(j)$ stands for the set of
$(n,k)$ clusters based at site $j$ (recall that $k$ indicates the number of
holes).
Then we have:
$$ h_4^L=2 f_{2,0}(1),\eq$$
$$ h_6^L=2 f_{4,0}(1)+4 f_{2,1}(1)-4 f_{2,0}(2) .\eq$$
One might hope that all the higher boundary terms could be similarly
expressed
in terms of the polynomials $f_{n,k}(j)$, which would significantly
reduce the number of coefficients in the ansatz for the boundary part of $H_n$.
It is easily seen to hold for the leading boundary term:
$$ h_{2m}^L= 2 f_{2m-2,0}(1) + {\rm lower\; orders} .\eq$$
Unfortunately, this is not true in general for the subleading terms
as can be seen from the exact expression for $h_8^L$
$$ \eqalign{
h_8^L &=2 f_{6,0}(1)-4 f_{4,0}(2) +2 f_{4,1}(1) + 8 f_{2,0}(3) +
2 f_{2,0}(2)\cr &+ 8 f_{2,1}(1) + 2 f_{2,2}(1)- 8 f_{2,1}(2)
+ 2 f_4(1,3,4,5) -2 f_4 (1,3,2,4). }  \eq$$
Since the last term corresponds to a {\it disordered}
 cluster, $h_8^L$ cannot be
rewritten as a linear combination of the simple polynomials
$f_n(\C)$ (with $\C$ ordered).
Higher-order boundary terms are rather complicated and no simple
pattern seems to emerge.


\newsec{The XY chain and related models}

In this section, we present the results for the
conservation laws of the open chain of the XY type. The basic open XY model
is defined by the hamiltonian:
$$H_{XY}=\sum_{j=1}^{N-1}  \a \s^x_j\s^x_{j+1}+ \b \s^y_j\s^y_{j+1}.
\eqlabel\hamxy$$
As is well known, by a the Jordan-Wigner transformation,
the XY chain (regardless of the boundary conditions)
can be reduced to a free fermionic theory.
Not surprisingly,  the  conserved charges for  free-fermion chains
have a particularly
simple form. For periodic boundary conditions, the conservation laws have been
described in [\ref{Gusev E V  {\it Theor. Math. Phys.}
{\bf 53} (1983), 1018}\refname\Gu,
\ref{Grady M {\it Phys. Rev} {\bf D25} (1982), 1103}\refname\Gra,
\ref{Araki H{\it Comm. Math. Phys. }{\bf 132} (1990),
155}\refname\Ara, \ref{Itoyama H and  Thacker H B {\it Nucl. Phys.} {\bf
B320}
(1989) 541}\refname\IT] (see also [\Ann]).

An interesting  feature of the closed
XY case is  the existence of  two independent families of  conservation laws,
which both persist
when the model is perturbed by a perpendicular (i.e.,  in the
$z$-direction) magnetic field. As we will see shortly,
when the closed chain is cut open, only half of each family survives.

We consider first the  particularly simple special case $\a=\b=1$ (the XX
model).
We introduce the notation
(for $n\ge 2$)
$$e_{n}^{\alpha\beta}(j)=
\s^\alpha_j\s^z_{j+1}\dots \s^z_{j+n-2}\s^\beta_{j+n-1},\eq$$
and
\eqn\hxypl{ \eqalign{ e_n^{+}(j) &=
 e_n^{xx}(j) +e_n^{yy}(j)\quad\quad n~{\rm even}, \cr &=
 e_n^{xy}(j) -e_n^{yx}(j)\quad\quad n~{\rm odd}, }}
\eqn\hxym{ \eqalign{ e^{-}_n(j) &=
 e_n^{xy}(j)-e_n^{yx}(j)\quad\quad n~{\rm even}, \cr &=
e_n^{xx}(j)+e_n^{yy}(j)\quad\quad n~{\rm odd},  }}
with $$ e_1^-(j) =-\s_j^z.\eq$$
For the periodic XX chain,  the two mutually commuting families of conserved
charges
are:
$$ ^cH_n^{\pm}=\sum_{j=1}^{N} e_n^{\pm}(j).\eq$$
In particular, there are two `two-spin hamiltonians':
$$^cH_2^{+}=\sum_{j\in\Lambda}
\s^x_j \s^x_{j+1}+ \s^y_j \s^y_{j+1} ,\eq $$
$$^cH_2^{-}=\sum_{j\in\Lambda} \s^x_j \s^y_{j+1}- \s^y_j \s^x_{j+1}. \eq $$
$^cH_2^{+}$ is invariant under global parity
transformations, i.e.,  $\s_j^a\to -\s_j^a$,  while
$^cH_n^{-}$ behaves like a pseudoscalar\foot{$^cH_2^{-}$ is a
special case of the Dzyaloshinski-Moriya interaction
[\ref{Dzyaloshinski I E {\it J. Phys. Chem. Solids} {\bf 4} (1958) 241;
Moriya T {\it Phys. Rev. Lett} {\bf 4} (1969), 228}].}.

The open chain versions of these hamiltonians
$$ H_{2}^{\pm}=\sum_{j=1}^{N-1} e_2 ^{(\pm)}(j) \eq$$
no longer commute with each other:
$$ [H_2^+, H_2^-] =4 i \s^z_N - 4i \s^z_1. \eq$$
To each hamiltonian, there corresponds a different infinite
family of conserved charges. Their bulk parts are given by
$$ H_n^{\pm; bulk}= \sum_{j=1}^{N-n+1} e_n^{\pm}(j). \eq$$
The family of conservation laws for the scalar
hamiltonian is:
$$ \eqalign{
H_n^{(1)}= & \; H_n^{+; bulk} +h_n^{+;L} + h_n^{+;R}\quad\quad n~{\rm even},
\cr = &
\; H_n^{-; bulk} +h_n^{-;L} + h_n^{-;R}\quad\quad n~{\rm odd},}\eq $$ where
$$ h_n^{\pm; L}=-\sum_{k=1}^{[n/2]-1} e_{n-2k}^{\pm} (k), \eq$$
$$ h_n^{\pm; R}=-\sum_{k=1}^{[n/2]-1} e_{n-2k}^{\pm} (N-k+1). \eq$$
For example,
$$ \eqalign{
h_3^{-;L}=& -e_1^-(1)=\s_1^z, \cr
h_8^{+;L}=& -(e_6^+(1) +e_4^+(2) +e_2^+(3) )
= -( \s^x_1 \s^z_2 \s^z_3 \s^z_4 \s^z_5 \s^x_6 +
\s^y_1 \s^z_2 \s^z_3 \s^z_4 \s^z_5 \s^y_6 \cr & +
 \s^x_2  \s^z_3 \s^z_4 \s^x_5 + \s^y_2 \s^z_3 \s^z_4 \s^y_5 +
 \s^x_3 \s^x_4 + \s^y_3 \s^y_4 ).} \eq$$
Obviously, $h_n^{\pm; R}$ can be obtained
from  $h_n^{\pm; L}$ via the chain reversal (\chainrev).
The conservation of the family $H_n^{(1)}$
$$[ H_2^+, H_n^{(1)}]=0 \eq$$
can be verified by a straightforward calculation, using
$$ [ H_2^+, e_n^\pm(j)]=\pm 2 i\; (-1)^n \left( e^\pm_{n+1}(j-1)
-e^\pm_{n+1}(j)  +e^\pm_{n-1}(j+1)-
e^\pm_{n-1}(j) \right) \eq$$
where $1<j<N$.

Similarly, one obtains the family of conservation laws
for the open chain pseudoscalar hamiltonian
$$\eqalign{
H_n^{(2)}= & H_n^{-; bulk} +g_n^{-;L} + g_n^{-;R}\quad\quad n~{\rm even}, \cr
=&
H_n^{+; bulk} +g_n^{+;L} + g_n^{+;R}\quad\quad n~{\rm odd},}\eq$$
where
$$ g_n^{\pm;L}=\sum_{k=1}^{[n/2]-1} (-1)^{k+1} e_{n-2k}^{\pm} (k) \eq $$
and $g_n^{\pm;R}$ is obtained from $g_n^{\pm;L}$ by chain reversal.
For example,
$$ \eqalign{ g_8^{-;L}=& e_6^-(1) + e_4^-(2)+ e_2^-(3) =
 \s^x_1 \s^z_2 \s^z_3 \s^z_4 \s^z_5 \s^y_6 - \s^y_1 \s^z_2 \s^z_3 \s^z_4 \s^z_5
\s^x_6
\cr &
- \s^x_2 \s^z_3 \s^z_4 \s^y_5 + \s^y_2 \s^z_3 \s^z_4 \s^x_5 +
\s^x_3 \s^y_4 - \s^y_3 \s^x_4 .} \eq$$

The two families $\{H_n^{(1)}\}$ and $\{H_n^{(2)}\}$ do not commute with each
other.
In particular, the charges $\{H_n^{(2)}\}$ do not commute with $H_2^+$.
Thus only half of the scalar and pseudoscalar families of the
periodic chain can be modified by adding the boundary terms so that they
still commute with the scalar hamiltonian of the
open chain.
Both families are invariant under a global spin
rotation around the $z$-axis:
it is easily checked  that all $H_n^{(i)}$'s commute with
the generator of such rotation, the $z$-component of the
total spin,
$$S^z=\sum_{j=1}^N \s_j^z.\eq$$
Therefore, both families survive when the model is perturbed by a
magnetic field term $h S^z$.

Consider now the anisotropic case (XY model):
$$ H=\sum_{j=1}^{N-1} (\a \s^x_j \s^x_{j+1} + \b \s_j^y
\s_{j+1}^y).\eqlabel\HXY$$
For the periodic XY chain, there are also two distinct
families  of conservation laws, all commuting together. One sequence is
given  by
$$ ^cH_n=\sum_{j=1}^N h^{XY}_n(j) \eq$$
with
$$ h^{XY}_n(j)= \a e_n^{xx} (j) + \b e^{yy}_n(j) + \a e_{n-2}^{yy}(j)+
\b e_{n-2}^{xx}(j). \eq$$
This expression gives the bulk density of the conserved charge
$H_n$ for the open XY chain
$$ H_n=H_n^{bulk}+ h_n^L + h_n^R, \eqlabel\hnXY$$
$$ H_n^{bulk}=\sum_{j=1}^{N-n+1} h^{XY}_n(j).\eq$$
The border part is given by
$$ \eqalign {
h_n^L=& -\sum_{k=1}^{[n/2]-1}
\{ \a^{1-p_k}\b^{p_k} [ e^{xx}_{n-2k}(k)+ e^{xx}_{n-2k}(k-1)] \cr &+
\a^{p_k}\b^{1-p_k} [ e^{yy}_{n-2k}(k)+ e^{yy}_{n-2k}(k-1)] \}
} \eq$$
where
$$ p_k={\frac {1} {2}} (1 -(-1)^k ) \eq$$
with the convention
$$ e^{\alpha\beta}_m(0)=0 .\eq$$
For example
$$ \eqalign{ h_8^L=& -\b e_6^{xx}(1) - \a e_6^{yy}(1) -
\a e_4^{xx}(1) - \b e_4^{yy}(1) -\a e_4^{xx}(2)\cr & - \b e_4^{yy}(2)-
\b e_2^{xx}(3) - \b e_2^{xx}(2) -\a e_2^{yy}(3)- \a e_2^{yy}(2). }\eq$$
The commutativity of
(\hnXY) with the hamiltonian (\HXY) can be established by a
direct calculation.
The second  family of conservation laws
(containing the
pseudoscalar XX hamiltonian)
can be modified in a similar way to account for the anisotropic deformation.

In the presence of both the anisotropy and the magnetic field
i.e., for the open XYh model
$$ H=\sum_{j=1}^{N-1} \a \s^x_j \s^x_{j+1} +\b \s^y_j \s^y_{j+1} +
\sum_{j=1}^N h \s^z_j, \eq$$
except for the special cases $\a = \b $ (XXh model) and $\b=0$ (the Xh model,
equivalent to the Ising chain)
there exist no counterterms that could be added to the bulk part
of $H_3$ to account for the open boundary conditions.
The first nontrivial charge in the nondegenerate case
is $H_4$, with
$$ \eqalign{H_4^{bulk}=&\sum_{j=1}^{N-3}(\a e_4^{xx}(j)+\b e_4^{yy}(j) )+
\sum_{j=1}^{N-2}(\alpha_3 e_3^{xx}(j)+ \beta_3 e_3^{yy}(j) ) \cr  &+
\sum_{j=1}^{N-1}(\alpha_2 e_2^{xx}(j)+ \beta_2 e_2^{yy}(j) )+
\kappa \sum_{j=1}^{N}\s_j^z}
\eq$$
and
$$ h_4^L=-\b \s_1^x\s_2^x -\a \s_1^y\s_2^y - h (1+\a/\b+\b/\a) \s_1^z,\eq$$
where
$$ \alpha_3=-h(\a+2\b)/\b, \eq$$
$$ \alpha_2=-2\a - \a^2/\b +h^2/\a+h^2/\b +\a \kappa /h \eq$$
and $\beta_i=\alpha_i (\a \leftrightarrow \b)$ ($i=2,3$),  $\kappa$
being arbitrary.

We end this section with a remark on the open Hubbard model:
$$ H= -2\sum_{~s=\ua, \da}
 \sum_{j=1}^{N-1}[ (a^\dg_{j,s} a_{j+1,s} + a^\dg_{j+1,s} a_{j,s}) +
4U\sum_{j=1}^N (n_{j,\ua}-1/2)( n_{j,\da}-1/2),\eq$$
where
 $a^\dg_{j,s}$ and $ a_{j,s}$ are
fermionic creation and annihilation operator of an electron of spin $s$
at site $j$,
satisfying the anti-commutation relation:
$$a^\dg_{j,s} a_{k,s'} + a_{k,s'} a^\dg_{j,s} = \delta_{j,k}\delta_{s,s'},
\eqlabel \anticomf$$
$U$ is a coupling constant and
$$n_{j,s} = a^\dg_{j,s} a_{j,s}.\eq$$
The integrability of this system has been established by means of the Bethe
Ansatz
in [\ref{Martins M J and Fye R M {\it J. Stat. Phys.}
{\bf 64} (1991) 271}].
As is well known, this model may be
equivalently regarded as two coupled XX chains:
$$H = \sum_{j=1}^{N-1} [\s^x_j\s^x_{j+1}+\s^y_j\s^y_{j+1}+\t^x_j\t^x_{j+1}+
\t^y_j\t^y_{j+1}] +U\sum_{j=1}^N \s^z_j\t^z_{j},\eqlabel\hubham$$
(where $\s$ and $\t$ denote two independent sets of
Pauli matrices). The local conserved charges of the Hubbard model have
been investigated in [\Ann] for the case of periodic boundary conditions.
For free open boundaries,  the first nontrivial charge is found to be
$$ H_4=H_4^{bulk}+ h_4^L+ h_4^R, \eq$$ with
$H_4^{bulk}$  given by the restriction of the corresponding closed chain
expression (see [\Ann]), and
$$ h_4^L= - (
\s_1^x \s_2^x + \s_1^y \s_2^y +
\t_1^x \t_2^x + \t_1^y \t_2^y +
4 U \s_1^z \t_1^z). \eq$$
Unlike the XY case, no odd-order charges exist.


\newsec{Further generalizations and conclusions}

\vskip 0.3cm

Our initial objective for this work was to obtain a closed form expression for
the
conserved charges of the open XXX chain, that is, to find the finite open chain
deformation of the Catalan tree pattern of the conserved
charges for the infinite chain. This has not been achieved, the main reason
being
that the spin polynomial basis used in the description of the charges in the
closed
case is inadequate for open chains. This, however, could be seen only after
looking
at the third nontrivial conservation law ($H_8$). The difficulties in studying
the
open chain conservation laws  also come from the absence of a  boost operator;
there
is thus no recursive way of constructing the charges.

In a sense, the problem has
been solved half-way: we showed that the expressions for the charges split in
two
parts, a known bulk part and boundary corrections;  we showed further how these
corrections can be calculated systematically (and wrote down the leading
boundary
contribution).

Moreover, we have demonstrated that an open chain has
half as many local conserved charges as its closed version with an
identical number of sites.  When one cuts open a closed spin chain,
there appears an additional,
 symmetry (the chain reversal symmetry (\chainrev)), an immediate consequence
of
which being that all odd-order charges cease to be conserved. This is
reminiscent
of the reduction, by a factor 2, of the order of the two-dimensional conformal
group induced by a  boundary: in presence of a boundary, the holomorphic and
antiholomorphic Virasoro modes are no longer independent!

It should be stressed that
the basic structure
of the even-order conservation laws generated from the transfer matrix
-- i.e., the separation into the bulk and boundary parts ---
is a direct consequence of the construction
(\tmodef).\foot{As shown in sect. 2, all of the
conserved charges generated by the transfer matrix of an open chain
may be obtained in this way.  Moreoer, the chain reversal symmetry guarantees
that
our construction based on solving commutativity constraints for
a boundary part added to the bulk part of an open chain charge,
cannot succeed for odd orders. Therefore, our construction
provides no more charges then the transfer matrix formalism.
However, it is not immediately clear whether our approach (or the transfer
matrix construction) exhausts all the possible
conservation laws for open chain. The problem of completeness
of the family of conserved charges  generated by the transfer
matrix is a very difficult one and has not yet been solved for generic (closed
or
open) integrable chains; it is certainly beyond the scope of our
investigation.}
The structure (\chargeform) characterizes
therefore a large class of integrable chains with nontrivial boundary
conditions.
Moreover, in models where the transfer matrix does not generate the complete
family of conservation laws (in the XY model for instance), the additional
conserved charges are expected to exhibit the same separation into  bulk and
boundary pieces.

The method for the calculation of the explicit expressions of the
local conserved
charges in the open XXX model presented in sect. 3 can be applied to
more general situations in a straightforward way.
For example,
through a simple unitary transformation, the exact expressions of $H_4,\;
H_6$ and
$H_8$ can
be translated into conserved charges for the XXZ model with anisotropy
parameter $\Delta=-1$ [\JMP].  Similarly, by reinterpreting the vector product
and the polynomials $f_{n,k}(j)$ in
terms of a Lie algebra commutator, these expressions apply directly to
the $su(N)$
version of the XXX chain, as well as to the octonionic chain [\JMP].
All these chains are particular examples
of a general algebraic model defined in
ref. [\JMP],
characterized by three arbitrary constants $\kappa_1, \kappa_2, \kappa_3$.
For periodic boundary conditions
this model has been proved to be integrable
provided that $\kappa_1+\kappa_3=2 \kappa_2$.
The same condition assures also the integrability of this general model with
free open boundary conditions. In particular, the boundary term in
$H_4$ is
$$ h_4^L= (2 {\kappa_2}/{\kappa_1}) \; f_{2,0}(1) .\eq$$

Although we have considered mainly open chains with free boundary conditions,
we
could treat more general boundary conditions in exactly the same way.  Adding
boundary terms in
$H_2$ simply induces extra terms in the higher order charges. For instance, by
adding the boundary term $k\sigma^z_1$ to the hamiltonian of the open XXX
chain  we find the following extra terms
in $H_4$:
$$
  k\sigma_1^x\sigma_2^z\sigma_3^x
+ k\sigma_1^y\sigma_2^z\sigma_3^y
- k\sigma_1^x\sigma_2^x\sigma_3^z
- k\sigma_1^y\sigma_2^y\sigma_3^z
-k^2\sigma_1^x\sigma_2^x-k^2\sigma_1^y\sigma_2^y-
k\sigma_1^z +(3k-k^3)\sigma_1^z . \eq$$
Anisotropic versions of the XXX model can be studied in similar way; again,
the only difference is that the structure of conserved charges is more
complicated. For instance, the first nontrivial conserved charge for the
XYZ chain with free open boundaries is
$$ H_4^{XYZ}=H_4^{bulk} + h_4^L+h_4^R, \eq$$
with
$H_4^{bulk}$ obtained from the corresponding closed XYZ expression,
(given in [\Ann]),
and
$$ h_4^L= \a (\b^2+\c^2) \s_1^x \s_2^x+
\b (\a^2+\c^2) \s_1^y \s_2^y+
\c (\a^2+\b^2) \s_1^z \s_2^z .\eq$$

For all these generalizations of the spin-1/2 XXX model, local conserved
charges of odd degrees are also absent.\foot{One might ask whether there are
some boundary conditions such that the only charges present, including the
hamiltonian, are of odd degree.
However, this seems impossible: for example,  for the XXX model
with nonperiodic boundary conditions, the lowest order odd charges
$H_3^{bulk}$ and $H_5^{bulk}$ cannot be deformed in a way that preserves
their mutual commutativity.}
As already mentioned, this result is
encoded in the transfer matrix
formulation of the open chain models.  Notice that
when there are nontrivial boundary conditions, hence nontrivial matrices
$X_\pm$, the spin reversal transformation
$\sigma_n\leftrightarrow\sigma_{N+1-n}$
must be accompanied by the interchange of $K_+$ and $K_-$.
We expect that the same conclusion will hold true for higher-spin
chains and even
nonhomogeneous chains with different spin representations at different sites
(provide that the sequence of spaces $V_1, V_2, \cdots ,V_N$ is invariant under
the interchange of sites $n$ and $N+1-n$).

For the simpler XY model, we could obtain the exact expression for all the
charges.  Here again, there is no open chain deformation for half of the
closed chain conserved charges, and as for closed chains, half of the charges
cannot be obtained from the power expansion of the transfer matrix.

It is worth mentioning that for a particular choice of boundary terms the XXZ
chain is endowed with quantum algebra symmetry $U_q[su(2)]$ [\ref{
Pasquier V and Saleur H {\it Nucl. Phys.} {\bf B} 330 (1990) 523}].
Clearly, the corresponding
family of local conserved charges is then invariant under the action of this
quantum
algebra. However, this additional symmetry does not seem to lead to
a significant simplification in the pattern
of the conserved quantities.

In ref [\ref{Grabowski M P and Mathieu P {\it J. Phys A: Math. Gen}.
{\bf 28} (1995) 4777}\refname\itest],
we have proposed a simple integrability test for closed chains
with
short-range intereactions. This test essentially boils down to the existence of
a conserved charge $H_3$ (more precisely, we conjectured that the nonexistence
of
$H_3$ means nonintegrability). This test needs to be modified for open chains
in an
obvious way; here it is the existence of $H_4$ that should characterize the
class of integrable models.

\newsec{Appendix}

For the sake of completeness, we give the basic commutators for the XXX model
of the type $[\S_i \cdot \S_{i+1}, f_n(\cal C)]$  (see eq. (4.13) in [\Ann])
in the graphical notation of section 3. In the expressions below,
$\cdots$  stand for an arbitrary number of filled ($\bullet$) or open ($\circ$)
dots.

$$ \contract{\cir\bullet}\;\cdots \bul =   2i\;  \bul\bul\cdots \bul  \;
\eqlabel\rulone$$
$$ \contract{\bul\circ}\;\cdots \bul =   -2i\;  \bul\bul\cdots \bul  \;
\eqlabel\ruletwo$$
$$ \contract{\bul\bullet}\;\cdots\bul =   4i\;  \cir\bul\cdots\bul -
4i\; \bul\cir\cdots\bul  \;  \eqlabel\rulethree$$
$$
\bul\cdots\contract{\bul\bullet}\;\cdots \bul =
2i\;  \bul\cdots \cir\bul\cdots\bul -
2i\;  \bul\cdots \bul\cir\cdots\bul
  \;  \eqlabel\rulefour$$
$$ \bul\cdots\contract{\bul\circ}\;\cdots\bul=-2i \;\bul\cdots(\bul\bul)\cdots
\bul
\eqlabel\rulefive$$
In the last equation, the parentheses indicate a modification of the
usual nesting pattern, e.g. the sequence
$\bul\bul(\bul\bul)\bul\bul $, starting at site 1, say,  corresponds to
the polynomial $(((\S_1\times \S_2)\times(\S_3\times\S_4))\times\S_5)\cdot
\S_6$.
Further basic commutators may be obtained applying the chain reversal
symmetry to the rules (\rulone)-(\rulethree) and (\rulefive).
In particular, we have:
$$ \bul\cdots\contract{\bul\circ} =   -2i\;  \bul\cdots \bul\bul  \;  \eq$$
$$ \bul\cdots\contract{\cir\bullet} =   2i\;  \bul\cdots \bul\bul  \;  \eq$$
$$ \bul\cdots\contract{\bul\bullet}\; =   4i\;  \bul\cdots\bul\bul -
4i\; \bul\cdots\bul\cir  \;  \eq$$
$$ \bul\cdots\contract{\cir\bullet}\;\cdots\bul=2i \;\bul\cdots(\bul\bul)\cdots
\bul
\eq$$


\bigskip
\centerline{{References}}
\immediate\closeout\refs \vskip 0.5cm
  \message{References}\input references

\end